\documentclass[runningheads,a4paper]{llncs}

\usepackage{amssymb}
\usepackage{graphicx}
\usepackage{bm}

\newcommand{\keywords}[1]{\par\addvspace\baselineskip
\noindent\keywordname\enspace\ignorespaces#1}

\begin{document}

\mainmatter  

\title{Efficient Algorithms for \\ Universal Quantum Simulation}

\titlerunning{Efficient Algorithms for Universal Quantum Simulation}

%
\author{Barry C.\ Sanders%
\thanks{This project has been supported by NSERC, CIFAR, AITF, USARO, MITACS and PIMS, and I acknowledge numerous valuable discussions with Nathan Wiebe about these concepts.}}
\authorrunning{Barry C.\ Sanders}

\institute{Institute for Quantum Science and Technology,
	University of Calgary, \\Calgary, Alberta T3A 0E1, Canada\\
\url{http://www.iqis.org/people/peoplepage.php?id=4}}

\toctitle{Efficient Algorithms for University Quantum Simulation}
\tocauthor{B. C. Sanders}
\maketitle
\begin{abstract}
A universal quantum simulator would enable efficient simulation of quantum dynamics by implementing quantum-simulation algorithms on a quantum computer.
Specifically the quantum simulator would efficiently generate qubit-string states that closely
approximate physical states obtained from a broad class of dynamical evolutions.
I provide an overview of theoretical research into universal quantum simulators and the strategies for minimizing computational space and time costs.
Applications to simulating many-body quantum simulation and solving linear equations are discussed
\end{abstract}

\section{Introduction}

A quantum computer could allow some problems to be solved more efficiently,
by enabling efficient execution of quantum algorithms,
as compared to executing classical algorithms on a classical computer
that are inferior for those problems~\cite{DiV95}.
The ``classical computer'' refers to a computer that is built strictly 
according to the principles of classical physics
but more specifically is equivalent to a Turing machine~\cite{BV97}.
The subtle issues of a quantized computer operating over real rather than binary fields are not discussed here~\cite{AHS09}.
The study of ``quantum simulation'' focuses on simulating properties and dynamics
of quantum systems whether by classical or quantum computation, 
and the topic of ``efficient algorithms for quantum simulation''
focuses on quantum simulation problems that do not have efficient classical algorithms.

Let me clear about terminology employed here.
By the term simulation, I mean that certain pre-specified properties of the quantum system are accurately predicted by the simulation but not necessarily all properties.
Accuracy refers to each answer being no worse than some error tolerance~$\epsilon$.
For example one might wish to know the mean momentum, the standard deviation of the momentum,
and average energy. 
The simulation is successful if these quantities are accurately predicted by the simulator even if other irrelevant quantities are poorly predicted.
The term efficiency refers to the simulation yielding an accurate solution to the problem
with a resource (e.g., run-time and space usage) cost that increases no
faster than a polynomial function of the input bit string and of $1/\epsilon$.

Explicitly defining simulation is important
because various notions
of quantum simulation using quantum computers, 
either purpose-built or universal, with various terminology.
The term ``digital quantum simulator'' is sometimes employed to refer to a programmable quantum simulator, and the term ``analogue quantum simulator'' refers to a quantum system
designed to behave analogously to a the quantum system being studied~\cite{BN09},
and usually these terms are employed when error correction is not assumed
hence making these systems not scalable.
Analogue quantum simulation is sometimes called ``quantum emulation''~\cite{NAB+09}.
Our term ``universal quantum simulator'' is in concordance with ``digital quantum simulator'' provided that the latter uses a fault tolerant architecture as we assume simulation on a scalable quantum computer.

Quantum simulation can deal with non-relativistic single-particle quantum mechanics described by Schr\"{o}dinger's equation
\begin{equation}
	{\rm i}\frac{\rm d}{{\rm d}t}|\psi(t)\rangle=\hat{H}(t)|\psi(t)\rangle
\label{eq:Schr}
\end{equation}
with self-adjointness $\hat{H}=\hat{H}^\dagger$ implying unitary dynamics,
but self-adjointness is not necessary.
Alternatively simulation of relativistic quantum mechanics or 
many-body quantum dynamics~\cite{AL97}
or quantum field theories~\cite{JLP12} may be sought.
For simplicity we focus on the easiest case of single-body dynamics~(\ref{eq:Schr})
and thence to the many-body case.

After choosing the equation to be studied,
the question then arises as to which problem is to be solved.
Two possible problems include 
solving the state~$|\psi(t)\rangle$ over some time domain
or determining the spectrum of the Hamiltonian~$\hat{H}$.
Instead of finding the spectrum or some aspect of the spectrum
such as the smallest or largest spectral gap, the problem could
be about finding eigenvectors of~$\hat{H}$ such as the ground state.
For simulation purposes a natural question would be to estimate the
expectation values of some observable
\begin{equation}
	\langle\psi(t)|\hat{\mathcal{O}}|\psi(t)\rangle.
\label{eq:mean}
\end{equation}
Some of the problems discussed here could be tractable on a classical computer hence making quantum algorithms uninteresting;
other problems such as finding ground states could be intractable as well on a quantum computer~\cite{KKR06}.

\section{Algorithms and complexity for quantum simulation}

For algorithmic quantum simulation we are interested in those problems that are 
intractable on a classical computer yet tractable on a quantum computer.
We can rule out solving problems that are amenable to the usual 
classical methods such as the following~\cite{WBHS10}.
One approach is to diagonalize the Hamiltonian directly,
which is always possible in principle but,
as the problem size is polylogarithmic in dimension and diagonalization is
polynomially expensive with respect to dimension,
the cost of diagonalization is thus superpolynomially expensive hence is not efficient in general.

Another approach to quantum simulation is to integrate the dynamical equation, for example Schr\"{o}dinger's equation~(\ref{eq:Schr}), directly.
For example the Runge-Kutta technique is popular.
Alternatively the dynamics can be tackled by constructing the evolution operator
and using the Magnus, or Baker-Campbell-Hausdorff method, expansion.
Product formul\ae~ are valuable as a unitary evolution can be factorized into an
approximate product of unitary evolutions. Product formul\ae~ include 
the Forest-Ruth or symplectic integration, method,
and the Trotter-Suzuki expansion is also valuable,
especially for quantum simulation as we shall see.

Quantum Monte Carlo simulations include
stochastic Green functions techniques and 
variational, diffusion or path-integral Monte-Carlo methods.
Density matrix renormalization group techniques have become popular especially
for one-dimensional many-body systems with slowly increasing entanglement with respect
to the number of particles.

\begin{figure}
	\includegraphics[scale=.48]{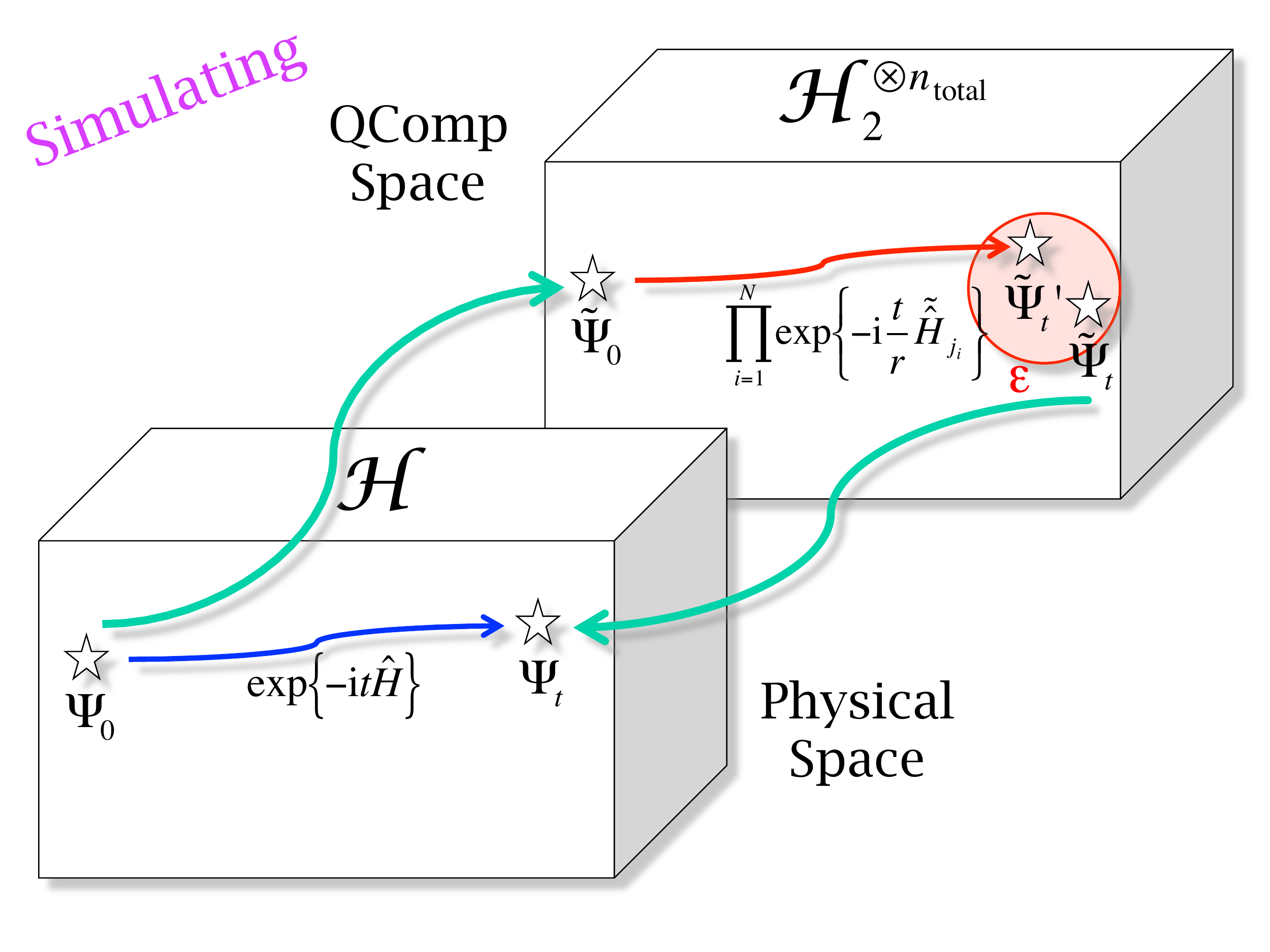}
\caption{
	Quantum simulation is depicted by showing evolution of a state~$\Psi_0$
	in a Hilbert space~$\mathcal{H}$ for the Physical Space and the
	evolution of the state's approximation $\tilde{\Psi}_0$ in the $n$-qubit space
	$\mathcal{H}_2^n$ in the quantum computer, or QComp Space.
	In the physical world, the evolution is given by $\exp\{-{\rm i}t\hat{H}\}$
	for~$\hat H$ the system Hamilttonian and~$t$ the time of evolution.
	The resultant state is~$\Psi_t$.
	In the quantum computer, all information is restricted to finite bit strings and all 
	quantum information to finite qubit strings so even continuous time~$t$
	is broken up into discrete intervals of duration~$t/r$ and the Hamiltonian
	matrix is approximated by $\tilde{\hat H}$.
	The resultant simulated state is~$\tilde{\Psi}'_t$, which is different from the 
	approximation of the true state~$\tilde{\Psi}_t$ by less than a distance~$\epsilon$.
	}
\label{fig:qsim}
\end{figure}

Perhaps the best insight into quantum simulation can be gained by studying
Feynman's own words in his seminal 1982 paper on quantum computing
based on a his keynote talk on the topic ``Simulating Physics With Computers''~\cite{Fey82}.
Feynman asks,
\begin{quote}
	Can a quantum system be probabilistically simulated
	by a classical (probabilistic, I'd assume) universal computer? 
	In other words, a computer which will give the same probabilities 
	as the quantum system does. 
	If you take the computer to be the classical kind I've described so far, 
	(not the quantum kind described in the last section) 
	and there're no changes in any laws,
	and there's no hocus-pocus, the answer is certainly, No!
	This is called the hidden-variable problem:
	it is impossible to represent the results of quantum mechanics
	with a classical universal device.
\end{quote}
The concept of quantum simulation can be understood from the schematic 
in Fig.~\ref{fig:qsim}.
The essence of this figure, which is fully explained in the caption,
is that the quantum simulation necessarily approximates all information and 
quantum information into bit strings and qubit strings and delivers an approximation
to the final state as a finite qubit string.

Let us now perform exegesis on Feynman's words to seek an understanding of what he meant.
In order to understand his meaning, we delve into computer science notions of complexity,
not something that Feynman himself used.
Thus, we seek to interpret a statement more than three decades old through the lens of 
modern computational complexity theory.

To understand, we cast quantum simulation as a decision problem:
the computational problem is constructed so that the answer can only be Yes or No.
To assess whether the quantum simulation is efficient,
the question is then how hard, i.e., how do the computational resources
scale with problem size expressed as the number of bits~$n$
required to specify the input state,
in order to answer the question?
Note that the resources to prove Yes or No can differ, which leads to complexity 
classes and their complements.
We are especially interested in the time and space costs,
which we denote as~$T$ and~$S$, respectively.

Quantum simulation problems are no worse then EXP,
which is the class of problems that can be solved with~$T$ and~$S$
increasing no more than an exponential function of~$n$.
That EXP is the worst case follows from using the Heisenberg matrix representation
for the dynamics and seeing that the size of the register and the computational
time for matrix operations leads to decision problems being in EXP.

Aaronson points out that Feynman (inadvertently) reduced the complexity of
quantum mechanics to PSPACE;
i.e., $S$ increases no more than polynomially in~$n$
by introducing path integrals~\cite{Aar13}.
The class of decision problems solvable efficiently on a quantum computer
is BQP, which refers to bounded-error quantum polynomial and is inside PSPACE.
The aim of quantum simulation thus needs to focus on narrower problems
than those in PSPACE.
Feynman's words ``give the same probabilities'' hints at the correct approach.
One should ask questions pertaining to expectation values of certain observables
and accept answers that are probabilistically equivalent to the true probabilities 
for these observables in the physical world.

Feynman's comment, ``classical kind \dots the answer is certainly, No!'' is more problematic.
He suggests that the classical simulation is provably inferior to the quantum simulation 
because of ``the hidden-variable problem:
it is impossible to represent the results of quantum mechanics
with a classical universal device''.
This question of provable superiority remains unresolved today,
and the hidden-variable problem does not lead to its resolution.
Feynman's idea that there is a strict separation between two computational
complexity classes can be regarded as a hard one to settle by 
thinking about this problem along the lines of any reduction in the 
polynomial complexity hierarchy. Such problems are famously difficult.

Lloyd recognized in 1996 that the key to formalizing Feynman's claim
lay in how to discrete the time evolution into discrete gate steps with a bound on 
the accumulated error due to time discretization~\cite{Llo96}.
Specifically Lloyd used the Trotter product formula
\begin{equation}
	{\rm e}^{{\rm i}t\left(\hat{A}+\hat{B}\right)}
		\to\lim_{n\to\infty}\left({\rm e}^{{\rm i}t\hat{A}/n}{\rm e}^{{\rm i}t\hat{B}/n}\right)^n.
\end{equation}
to approximate the evolution operator,
with the Hamiltonian expressed as the sum $\sum_{j=1}^m\hat{H}_j$ as
\begin{equation}
	\exp\left\{-{\rm i}t\sum_{j=1}^m\hat{H}_j\right\}
		=\left(\prod_{i=1}^N\exp\left\{-{\rm i}\frac{t}{r}\hat{H}_{j_i}\right\}\right)^r
			+\sum_{j>j'}\left[\hat{H}_j,\hat{H}_{j'}\right]\frac{t^2}{2r}+{\rm error}.
\end{equation}
Lloyd proved that this simulation had a $T$ and $S$ costs that are only poly$(n)$.
This result can generalized to a time-dependent Hamiltonian 
and the errors tightened~\cite{WBHS10}.

In 2003, Aharonov and Ta-Shma analyzed the general question of what Hamiltonian 
systems are efficiently simulatable~\cite{AT03}.
Their work was motivated by strong claims about adiabatic quantum computing 
solving NP-Hard problems.
They tackled the problem by considering which quantum states can be efficiently generated
and cast the problem into the oracle setting: $\hat{H}$ is in a black-box, which is queried
with an assigned cost per query.
A key result of their work is their demonstration of 
equivalence between quantum state generation and statistical zero knowledge problems.
Another important result is the Sparse Hamiltonian Lemma:
If $\hat{H}$ acting on $n$ qubits is $d$-sparse
s.t.\ $d\in O({\rm poly}n)$ and the list of nonzero entries in each row is efficiently computable,
then $\hat{H}$ is \emph{simulatable} if $\|\hat{H}\|\leq{\rm poly}n$.

\begin{figure}
	\includegraphics[width=\linewidth]{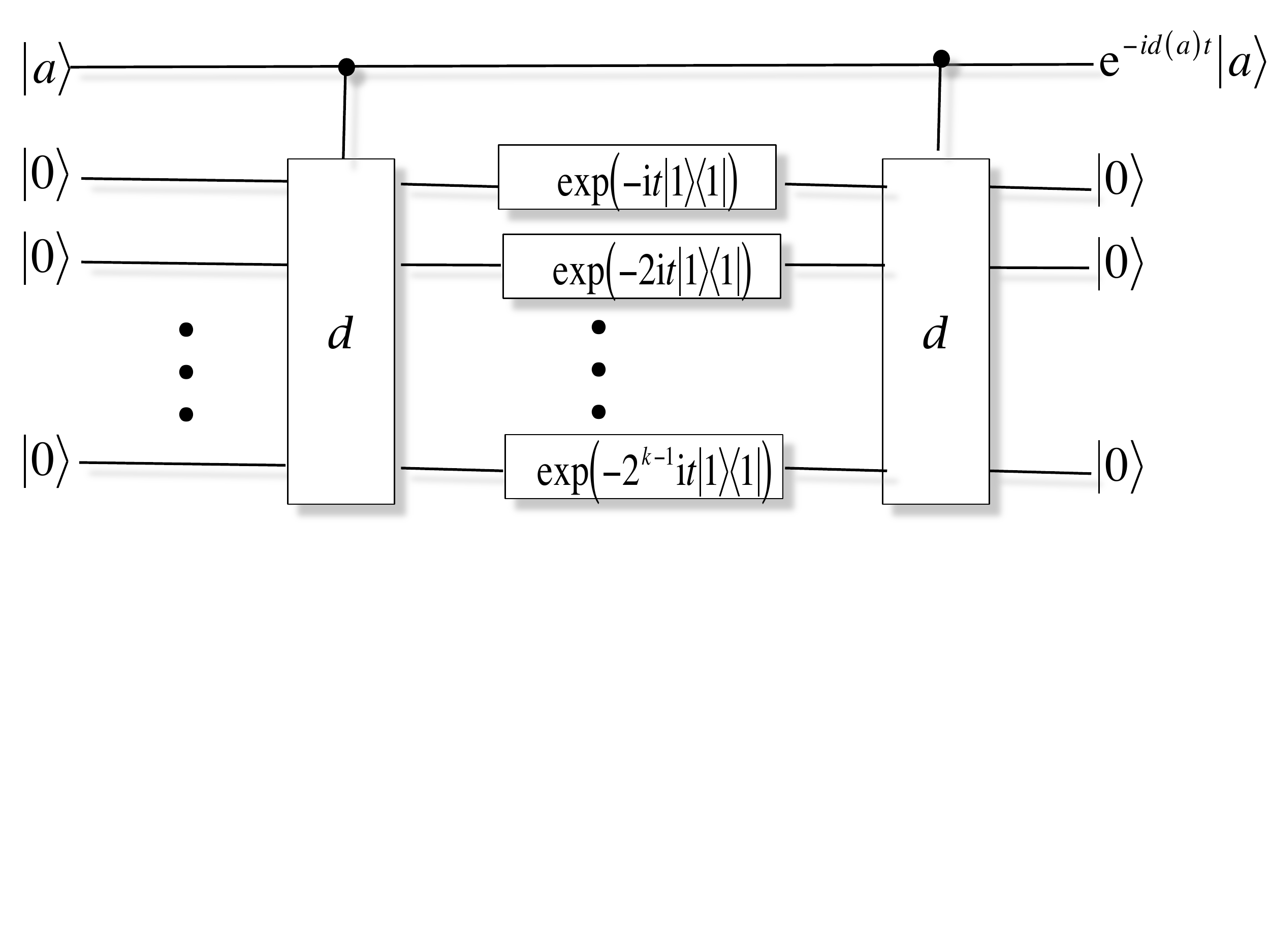}
\caption{
	Simulating evolution for diagonal~$\hat{H}$ with
	$d(a)=\langle a|\hat{H}|a\rangle\in\{0,1\}^k$.
	The row numbers~$a$ of the Hamiltonian are written onto a string
	of qubits, and the string of qubits in the $|0\rangle$ state are ancillary.
	The depicted circuit circuit then effects the transformation
	$|a,0\rangle\mapsto|a,d(a)\rangle\mapsto{\rm e}^{-{\rm i}td(a)}|a,d(a)\rangle
		\mapsto{\rm e}^{-{\rm i}td(a)}|a,0\rangle={\rm e}^{-{\rm i}\hat{H}t}|a,0\rangle$.
	}
\label{fig:diag}
\end{figure}

We can use Childs's rules for simulatability~\cite{Chi04}
to augment the Sparse Hamiltonian Lemma.
The system is simulatable if the Hamiltonian is a sum
$\sum_i\hat{H}_j$ with each $\hat{H}_j$ acting on $O(1)$ qubits or
is $\sqrt{-1}\times$ a commutator of two simulatable Hamiltonians or
is efficiently convertible to a simulatable Hamiltonian by efficient unitary conjugation
or is sparse and efficiently computable.
The basic element for simulating Hamiltonian evolution is depicted in Fig.~\ref{fig:diag}
for the case of a diagonal Hamiltonian.
The circuit is easily generalized to one-sparse Hamiltonian generated evolution
whether diagonal or not~\cite{CCD+03}.

The quantum simulation circuit is designed to approximate the desired 
unitary evolution operator~$U$ by a sequence 
$\prod_{\nu=1}^NU_{j_\nu}$ where each~$U_{j_\nu}$
is generated by one of~$m$ one-sparse Hamiltonians.
Generalizing the Trotter formula using the Suzuki iteration method
leads to a much more efficient way of performing
this unitary factorization,
i.e., to a product of unitary gates with the 
length of this sequence of unitary gates being $t^{1+o(1)}$~\cite{BACS07a,BACS07b}.

The Hamiltonian in the oracle is promised to be $d$-sparse with
$d\in$poly$(n)$.
This creates the algorithmic challenge of reducing the $d$-sparse Hamiltonian into
a disjoint sum of one-sparse Hamiltonians.
The decomposition is aided by first converting sparse Hamiltonians into 
graphs of low degree and then colouring the graph so that it is a disjoint union
of degree-one graphs; hence the corresponding Hamiltonian is a direct
sum of one-sparse Hamiltonians each corresponding to one colour of the 
graph~\cite{BACS07a,BACS07b}.

The Hamiltonian is converted to a graph as follows.
Let~$x$ label a row of the Hamiltonian matrix and~$y$ the column.
As the Hamiltonian is $d$-sparse,
there are at most~$d$ column numbers that hold nonzero elements for row~$x$.
We call these column entries $y_{1,\dots,d}$ in no particular order;
i.e., the increasing sequence of indices $1,\dots,d$ does not imply increasing values
of~$y_i$.
Now construct the graph by assigning each~$x$ a vertex so that there are now
$2^n$ vertices but no edges yet.

For given~$x$, we construct an edge to another vertex value~$x'$ if
$x'=y_i$ such that $y_i$ is one of the column indices where row~$x$
and column~$y_i$ has a nonzero Hamiltonian matrix element.
The weight of the edge is the value of that matrix element $\langle x|\hat{H}|y_i\rangle$.
If we simplify~$\hat H$ to having only real matrix entries and note that $\hat{H}=\hat{H}^\dagger$,
then we can assign the Hamiltonian an undirected graph because 
$\langle x|\hat{H}|y_i\rangle=\langle y_i|\hat{H}|x\rangle$.
The Hamiltonian is thus faithfully represented by a degree-$d$ undirected graph.
A superior colouring algorithm that yields a direct sum of one-sparse
Hamiltonians can reduce $T$ and $S$ costs
for the associated quantum query algorithm to determine the sequence of
operations for evolution generated by a $d$-sparse Hamiltonian.

Table~\ref{table:costs} provides a summary of some advances over the years in reducing $T$ and $S$ costs. In some cases one cost is reduced at the expense of the other.
Although the efficiency of quantum simulation has been known for quite some time,
quantum simulations could be the first practical application of quantum computing.
Cost reductions reduce the waiting time for non-trivial quantum simulations to become
a reality and hence are important.

\begin{table}
\begin{tabular}{|l|l|l|l|}
		\hline
	Who&Year&T&S\\ \hline
	Lloyd\cite{Llo96}&1996&$O(t^2)$&$O(n)$\\ \hline
	AT\cite{AT03}&2003&$O\left(n^9d^4\frac{t^2}{\epsilon}\right)$&$O(n)$\\ \hline
	Childs\cite{Chi04}&2003
		&$O\left(n^2d^{4+o(1)}\frac{t^{3/2}}{\sqrt{\epsilon}}\right)$&$O\left(n\right)$\\ \hline
	BACS\cite{BACS07a}&2007
		&$O\left(\log^*\!nd^{4+o(1)}
			\frac{t^{1+1/2k}}{\epsilon^{1/2k}}\right)$&$O(n\log^*\!n)$\\ \hline
	CK\cite{CK11}
		&2010&$O\left(\left[d^3+d^2\log^*\!n\right]
			\frac{t^{1+1/2k}}{\epsilon^{1/2k}}\right)$
		&$O\left(nd+n\log^*\!n\right)$\\ \hline
	CB\cite{CB12}&2010
		&$O\left(\|\hat{H}\|_{\rm max}d\frac{t}{\sqrt{\epsilon}}\right)$& $\bullet$ \\ \hline
\end{tabular}
\caption{
	Key developments in reducing time $T$ and space $S$ costs for a quantum computer
	to simulate time-independent Hamiltonian generated evolution
	as a function of the number of qubits~$n$ representing the system,
	the sparseness~$d$ of the Hamiltonian,
	the allowed error~$\epsilon$,
	and the norm of the Hamiltonian $\|\hat{H}\|$.
	The authors are listed in the first column along with references and the year in the
	second column. The final row and column is given by $\bullet$ to show that the 
	space cost is not explicitly known.
	The iterated logarithm $\log^*$ in the table is the number of successive
	iterations of the base-two logarithm function required to reduce the number
	to one or less.
	}
\label{table:costs}
\end{table}

\section{Applications}
Although quantum computing was founded on the principle of quantum 
simulation, other algorithms such as factorization have dominated the field for many
years. The reason quantum simulation is back in full force can be understood
from the prescient quote from a 1997 paper by Abrams and Lloyd\cite{AL97}:
\begin{quote}
	But the problem of simulation --- that is, the problem of 
	modeling the full time evolution of an arbitrary quantum system --- 
	is less technologically demanding.
	While thousands of qubits and billions of quantum logic operations are needed to solve 
	classical difficult factoring problems [16],
	it would be possible to use a quantum computer with only a few tens of qubits
	and a few thousand operations to perform simulations
	that would be classical intractable [17].
\end{quote}
Abrams and Lloyd specifically showed that the quantum simulator would
efficiently simulate fermionic systems.
Combined with other results on bosonic and anionic systems,
the quantum simulator is thus known to be an efficient simulator of all types of
many-body systems.

Various many-body systems are considered for experimental quantum simulation
in order to learn properties about the system that are unreachable with classical simulations due to intractability.
Let us assign $X$, $Y$ and $Z$ as the Pauli operators on a single qubit.
The Hamiltonians for these many-body systems include the Ising Hamiltonian
$J\sum_{\langle i,j\rangle}Z_i\otimes Z_j+B\sum_iX_i$,
the XY Hamiltonian
$J_x\sum_{\langle i,j\rangle}X_i\otimes X_j+J_y\sum_{\langle i,j\rangle}Y_i\otimes Y_j$,
the Heisenberg Hamiltonian
$J_x\sum_{\langle i,j\rangle}X_i\otimes X_j+J_y\sum_{\langle i,j\rangle}Y_i\otimes Y_j$
and the honeycomb Hamiltonian
$J_x\sum_{x{\rm -link}}X_i\otimes X_j-J_y\sum_{y{\rm -link}}Y_i\otimes Y_j
		-J_z\sum_{x{\rm -link}}Z_i\otimes Z_j$.
Whereas earlier the algorithm for simulation is designed for the broadest
class of simulatable Hamiltonians,
if the Hamitlonian is known explicitly and is a sum of strictly local Hamiltonians,
then there is a straightforward circuit-construction algorithm for unitary gates generated by a tensor product of Pauli operators~\cite{RWS12}. 

Whereas quantum simulators are evidently useful for simulating quantum dynamics
by design, they can be used more broadly, for example to solve
giant sets of coupled linear equations~\cite{HHL09}.
This approach takes quantum simulators beyond applicability just
to quantum systems, but we have to be careful about what 
we mean by ``solve'' as we had to be careful about what we meant by
``solve'' Schr\"{o}dinger's equation earlier.

The problem to be solved by the quantum linear equation solver can be understood 
by the following statement.
\begin{quote}
	Given matrix~$A$ vector~$\bm{b}$, and matrix~$M$,
	find a good approximation of~$\bm{x}^{\rm T}M\bm{x}$
	such that $A\bm{x}=\bm{b}$.
\end{quote}
The strategy for using a quantum simulator to solve this problem is as follows.
Begin by replacing $\bm b$ by the quantum state
$|\bm{b}\rangle=\sum_{i=1}^Nb_i|i\rangle$
with~$|i\rangle$ the computational basis.

The solution would be $|\bm{x}\rangle=\hat{A}^{-1}|\bm{b}\rangle$,
but inverting $\hat{A}$ is hard so a method has to be found to circumvent this difficulty.
The operator $\hat{A}$ has eigenvalues~$\lambda_j$
and eigenvectors~$|u_j\rangle$ for $j=1,\ldots,N$,
and we express $|\bm{b}\rangle=\sum_{j=1}^N\beta_j|u_j\rangle$
in the $\hat{A}$-eigenbasis.
The concept is to recognize that
\begin{equation}
	|\bm{x}\rangle=\hat{A}^{-1}|\bm{b}\rangle
		=\sum_{j=1}^N\frac{\beta_j}{\lambda_j}|u_j\rangle.
\end{equation}
This approach is achieved by using	the phase-estimation approach,
namely by taking $\bm{b}\rangle$ with ancilla to obtain
$\sum_{j=1}^N\beta_j|u_j\rangle|\lambda_j\rangle$.
Then the non-unitary linear map
$|\lambda_j\rangle\mapsto\lambda_j^{-1}|\lambda_j\rangle$
is constructed in a quantum circuit.
Finally the circuit uncomputes $|\lambda_j\rangle$ to obtain
the approximation~$|\bm{x}\rangle$.

\section{Conclusions}
This article provides an overview of algorithmic quantum simulation,
approaches to implementing and improving these algorithms,
and applications of quantum algorithms for quantum simulation.
Theoretical research in this area is challenging because it draws in so
many different techniques from such different areas, for example
graph theory, operator algebra, and computational complexity.
The field is exciting from a technological perspective because
non-trivial problems could be solved with smaller quantum computers 
than for other planned applications of quantum computing such as to factorization.

\end{document}